# New clustering approach for symbolic polygonal data: application to the clustering of entrepreneurial regimes

Andrej Srakar

Institute for Economic Research (IER) and School of Economics and Business, University of Ljubljana, Slovenia, srakara@ier.si

Marilena Vecco

Burgundy School of Business – Université Bourgogne Franche-Comté, France, marilena.vecc@bsb-education.com

**Abstract**

Entrepreneurial regimes are topic, receiving ever more research attention. Existing studies on entrepreneurial regimes mainly use common methods from multivariate analysis and some type of institutional related analysis. In our analysis, the entrepreneurial regimes is analyzed by applying a novel polygonal symbolic data cluster analysis approach. Considering the diversity of data structures in Symbolic Data Analysis (SDA), interval-valued data is the most popular. Yet, this approach requires assuming equidistribution hypothesis. We use a novel polygonal cluster analysis approach to address this limitation with additional advantages: to store more information, to significantly reduce large data sets preserving the classical variability through polygon radius, and to open new possibilities in symbolic data analysis. We construct a dynamic cluster analysis algorithm for this type of data with proving main theorems and lemmata to justify its usage. In the empirical part we use dataset of Global Entrepreneurship Monitor (GEM) for year 2015 to construct typologies of countries based on responses to main entrepreneurial questions. The article presents a novel approach to clustering in statistical theory (with novel type of variables never accounted for) and application to a pressing issue in entrepreneurship with novel results.

**Keywords:** symbolic data; polygonal cluster analysis; entrepreneurial regimes

1. Introduction

Several different classifications of EU countries based on welfare regimes, type of capitalism and other institutional characteristics have been proposed in the recent decades. In their formulation, Hall and Soskice (2001) identify a core distinction between two types of political economies: liberal market economies, in which firms coordinate their activities primarily via firm hierarchies and competitive market arrangements, and coordinated market economies, in which coordination relies more heavily on non-market relationships (see Dilli and Elert, 2016). Esping-Andersen's (1990) original classification is composed of three models: Liberal, Social-Democratic and Continental, with later studies pointing to existence of Mediterranean and Eastern European regimes. In their analysis, Dilli and Elert (2016) derive six groups of entrepreneurial regimes, including a separate cluster for the Eastern European countries.

Entrepreneurial activities highly differ across countries. This holds true for different measures of entrepreneurship such as start-up activity, business ownership, small business share, nascent entrepreneurship and the preference and motives for entrepreneurship. Besides individual characteristics (risk tolerance, entrepreneurial culture, etc.), the level of economic development and cultural aspects are often mentioned as the principal drivers of entrepreneurial activities.

Although there are different streams and approaches of the institutional theory within the realm of entrepreneurship, the approach introduced by North (1994) is still the most used. This author defined institutions as »rules of game in the society« and constraints that structure human interaction … made up of formal constraints (for example, rules, laws, and constitutions) (North, 1994, p. 360). North classified the formal and informal institutions impacting organisations and organisational actors into regulatory, normative and cognitive categories.

Furthermore, some scholars observed, relevant cross-national differences may be embedded in historical experiences, institutional heritage, norms, or cultural values. These differences can provide idiosyncratic institutional milieus for entrepreneurial activities.

As noted by scholars, the set of regulatory procedures and administrative constraints may negatively impact the entrepreneurial activity as entrepreneurs have to spend extra time and resources to fit in the administrative system instead of being devoted to develop their business content. Furthermore, quality of institutions is regarded as many scholars to have a significant impact on the modelling of entrepreneurial regimes (for more see e.g. Dilli and Elert, 2016).

## 2. Polygonal variables in symbolic data analysis

Due to the nature of GEM database, which is a large scale database, including for each year more than 100,000 respondents, we utilise a symbolic data analysis, which is a special type of statistical analysis of large datasets, developed in recent decades (see e.g. Billard and Diday, 2003; 2006; Diday and Noirhomme-Fraiture, 2008; Noirhomme-Fraiture and Brito, 2011). In the classical data framework one numerical value or one category is associated with each individual (microdata). However, the interest of many studies lays in groups of records gathered according to the characteristics of the individuals or classes of individuals, leading to macro data. The traditional approach for such kind of situations is to associate with each individual or class of individuals a central measure, e.g., the mean or the mode of the corresponding records. However, with this option the variability across the records is lost. To avoid this unsatisfactory result, Symbolic Data Analysis (SDA) proposes that a distribution or an interval of the individual records' values is associated with each unit, thereby considering new variable types, named symbolic variables. One such type of symbolic variable is the histogram-valued variable, where to each entity under analysis corresponds an empirical distribution that can be represented by a histogram or a quantile function. To this purpose, it is necessary to adapt concepts and methods of classical statistics to new kinds of variables. Furthermore, our analysis derives from symbolic polygonal variable concept, differing in several aspects from polygonal spatial clustering in Joshi (2011).

A *multi-valued* symbolic variable $Y$ is one whose possible value takes one or more values from the list of values in its domain $\mathcal{Y}$. An *interval* symbolic variable $Y$ is one whose possible value takes values in an interval, i.e. $Y = \xi = [a, b] \subset \mathbb{R}$, with $a \leq b$, $a, b \in \mathbb{R}$. Let the random variable $Y$ takes possible values $\{\eta_k, k = 1,2, ...\}$ over a domain $\mathcal{Y}$. Then, a particular outcome is *modal valued* if it takes the form $Y(\omega_u) = \xi_u = \{\eta_k, \pi_k; k = 1, ..., s_u\}$ for an observation $u$ where $\pi_k$ is a non-negative measure associated with $\eta_k$ and where $s_u$ is the number of values actually taken from $\mathcal{Y}$.

Many methods have been developed for cluster analysis of interval data. Methods based on dissimilarities generally use adaptations of K-means (De Souza and De Carvalho, 2004; De Carvalho, Brito and Bock, 2006; Chavent et al., 2006). Alternative approaches propose suitable dissimilarity measures for interval data, and then use the K-means algorithm to obtain a partition that locally optimizes a criterion measuring the fit between the cluster composition and their prototypes.

Fuzzy K-means methods for interval data generally result from adapting the classical fuzzy c-means algorithm, using appropriate distances, as is done for the crisp algorithms. Other extensions use adaptive distances or multiple dissimilarity matrices. Hierarchical or pyramidal clustering has been used by Brito (1994; 1995). A monothetic clustering method using a divisive approach has been used in Chavent (1998) who uses a criterion that measures intraclass dispersion using distances appropriate to interval-valued variables. The algorithm successively splits one cluster into two sub-clusters, according to a condition expressed as a binary question on the values of one variable. Kohonen maps have been used by Bock (2002; 2008). Different dynamic algorithms have been applied as well, in particular in Chavent et al. (2006).

A novel type of symbolic variables, closely related and deriving from interval ones, have been proposed in a recent article of Silva and colleagues (2019). They define polytope as a convex hull of a compact non-empty finite set. In general, the face structures of a convex polytope are significantly more simple than convex hull. A polytope $P = conv\{x_1, \ldots, x_l\}$ is called a $k$-polytope if $dim P = k$. This means that some $(k+1)$ subfamily of $(x_1, \ldots, x_l)$ is affinely independent, but $(k+2)$ is not affinely independent. In the following we assume the same number of vertices for all variables.

**Definition 1:** Let $\Omega$ be polygons space and let $Z$ be a random variable such that $Z: \Omega \to \mathbb{R}^2$. This random variable assumes values in polygon $(P)$ with $L$ vertices, then $Z = \xi = \{(a_1, b_1), \ldots, (a_L, b_L)\} \subset \mathbb{R}^2$. It can also be rewritten by $Z = \xi = (\xi_1, \xi_2)$, where $\xi_1 = \{a_1, \ldots, a_L\}$ and $\xi_2 = \{b_1, \ldots, b_L\}$.

**Definition 2:** Let $n_z$ be the number of individuals in a class $z$. Each individual is described by a continuous variable $X$. A polygon $P_z$ with $l$ vertices for $l \leq n_z$ inscribed in circumference can be obtained by

$$P_{zl} = (a_{zl}, b_{zl}) = \left(c_z + r_z \cos\left(\frac{2\pi l}{L}\right), c_z + r_z \sin\left(\frac{2\pi l}{L}\right)\right) \quad (1)$$

where $c_z$ is the center of the polygon $z$ (mean of $X$ in class $z$) and $r_z = 2 \times sd(\chi_z)$ is the radius of the polygon (or circumference) $z$ where $sd(\chi_z)$ is the standard deviation of $X$ in class $z$, respectively. $P_{zl}$ represents a vertex of the polygon $P_z$ and $l = 1,2,\ldots,L$ for $L \in \mathbb{N} \geq 3$ is the number of vertices of this polygon.

One of main advantages of using polygonal data is relaxation of the equidistribution hypothesis, inherent (and problematic) for interval data.

Polygonal equidistribution hypothesis (which is a generalization of interval-related equidistribution) can be expressed as:
1. Each observation $u \in S$ is equiprobable, i.e., each observation is selected with probability $1/m$ where $m$ is the cardinality of sample space $(S)$;
2. We define $Z_u$ in the polygon for each $u \in S$, and $Z_u$ has uniform distribution in the polygon.
   Empirical pdf in any polygon:

$$f_Z(\xi) = \begin{cases} \dfrac{1}{m}\sum_{u\in S}\dfrac{1}{A_u} & if\ \xi \in P \\ 0 & otherwise \end{cases}$$

where

$$A_u = \frac{1}{2}\sum_{u\in S}|\sum_{i=1}^{L} b_{u,i}(a_{u,i+1} - a_{u,i-1})|$$

$A_u$ is the area of the polygon $P_u$ calculated by simple shoelace's equation (e.g. Zwillinger, 2011) for all $P_u \in S$ and it assumes $L$ vertices.

### 3. Novel clustering algorithm for polygonal data

We propose a novel and, actually, the first clustering algorithm for data described above, polygonal data/variables. It is based on combination of Hausdorff and City-Block distance, as defined below.

**Definition 3:** We define the distance between two polygons $p_1^z$ and $p_2^z$ as a combination of Hausdorff and City-Block distance:

$$d_H(p_1^z, p_2^z) = \max_z(|a_1^z - a_2^z| + |b_1^z - b_2^z|) \qquad (2)$$

where the maximization takes place over all correspondent vertices of the two polygons, $a_1^z$ and $b_1^z$ are coordinates of the corresponding vertex of polygon $p_1^z$, while $a_2^z$ and $b_2^z$ are coordinates of the corresponding vertex of polygon $p_2^z$.

Cumulative distance (distance criterion) between set (matrix) of polygons is defined accordingly as:

$$d_{TH}(p) = \sum_{z=1}^{p} \max_z(|a_1^z - a_2^z| + |b_1^z - b_2^z|) \qquad (3)$$

where the summation runs over the full matrix of $p$ polygons.

We also define the following:

**Definition 4:** The prototype $G = (g^1, \dots, g^p)$ of a cluster $C$ is a matrix of $p$ polygons which minimizes the adequacy criterion:

$$f_C(G) = \sum_{s\in C} d_{TH}(x_s, G) = \sum_{s\in C}\sum_{z=1}^{p} d_H(x_s^z, g^z) = \sum_{s\in C} \tilde{f}_C(g^z) \qquad (4)$$

Our main theoretical result is theorem below.

**Theorem 1:** *Derivation of prototype $G$ is equivalent to solving two separate minimization problems like in Chavent et al. (2006):*

$$\min \sum_{s\in C}|\mu^z - c_s^z| \qquad (5)$$

$$\min \sum_{s\in C}|\lambda^z - r_s^z| \qquad (6)$$

*The solutions $\hat{\mu}^z$ and $\hat{\lambda}^z$ are respectively the median of the set $\{c_z^s, s \in C\}$ of the polygon centers, and the median of the set $\{r_z^s, s \in C\}$ of their radiuses.*

**Proof:**

The problem of deriving the prototype in Definition 3 is equivalent to finding the polygon $g^z$ for $(z = 1, ..., p)$ which minimizes:

$$\tilde{f}_C(g^z) = \sum_{z=1}^{p} d_H(x_s^z, g^z) = \sum_{z=1}^{p} \max_z(|g_a^z - a_i^z| + |g_b^z - b_i^z|) \qquad (7)$$

If we insert the respective coordinates for prototype and each respective polygon vertex into (7), the equation for the Hausdorff/City-Block distance is transformed into:

$$d_H(x_s^z, g^z) = \max_z(|\hat{\mu}^z - c_i^z + (\hat{\lambda}^z - r_i^z)\cos\frac{2\pi l}{L}| + |\hat{\mu}^z - c_i^z + (\hat{\lambda}^z - r_i^z)\sin\frac{2\pi l}{L}|)$$

$$\leq \max_z(|\hat{\mu}^z - c_i^z| + |(\hat{\lambda}^z - r_i^z)\cos\frac{2\pi l}{L}| + |\hat{\mu}^z - c_i^z| + |(\hat{\lambda}^z - r_i^z)\sin\frac{2\pi l}{L}|)$$

$$= \max_z(2|\hat{\mu}^z - c_i^z| + |(\hat{\lambda}^z - r_i^z)\cos\frac{2\pi l}{L}| + |(\hat{\lambda}^z - r_i^z)\sin\frac{2\pi l}{L}|)$$

$$= \max_z(2|\hat{\mu}^z - c_i^z|) + \max_z(|(\hat{\lambda}^z - r_i^z)\cos\frac{2\pi l}{L}| + |(\hat{\lambda}^z - r_i^z)\sin\frac{2\pi l}{L}|)$$

$$= \max_z(2|\hat{\mu}^z - c_i^z|) + \max_z(|(\hat{\lambda}^z - r_i^z)||\cos\frac{2\pi l}{L}| + |(\hat{\lambda}^z - r_i^z)||\sin\frac{2\pi l}{L}|)$$

$$= \max_z(2|\hat{\mu}^z - c_i^z|) + \max_z(|(\hat{\lambda}^z - r_i^z)|(|\cos\frac{2\pi l}{L}| + |\sin\frac{2\pi l}{L}|)) \qquad (8)$$

where the maximization runs over all $L$ vertices of the polygon $z$. The equality in the second line inequality relationship is satisfied iff $\hat{\mu}^z - c_i^z$ and $(\hat{\lambda}^z - r_i^z)\cos\frac{2\pi l}{L}$ respectively $\hat{\mu}^z - c_i^z$ and $(\hat{\lambda}^z - r_i^z)\sin\frac{2\pi l}{L}$ are of the same sign (positive or negative).

As the maximization runs over the vertices, it is clear that expressions $\hat{\mu}^z - c_i^z$ and $\hat{\lambda}^z - r_i^z$ are fixed for each polygon. It follows from basic trigonometry that the same sign is achieved if $0 \leq \frac{2\pi l}{L} \leq \frac{\pi}{2}$ or $\pi \leq \frac{2\pi l}{L} \leq \frac{3\pi}{2}$. Furthermore, exact maximum of $\sqrt{2}$ of the expression $|\cos\frac{2\pi l}{L}| + |\sin\frac{2\pi l}{L}|$ is achieved for $\frac{2\pi l}{L} = \frac{\pi}{4} + 2k\pi, k \in \mathbb{Z}$. It is important to repeat this maximum does not depend on expressions $\hat{\mu}^z - c_i^z$ and $\hat{\lambda}^z - r_i^z$ which are fixed for each polygon.

The expression in (7) optimizes the adequacy criterion from Definition 3 over a set of polygons in cluster $C$. Hence, the exact value of maxima in (8) which depends only on specific polygon (and is fixed for each polygon) is not important. But this means the optimization in (7) is dependent only upon (separate) minimization of expressions $\min\sum_{s\in C}|\mu^z - c_s^z|$ and $\min\sum_{s\in C}|\lambda^z - r_s^z|$. This is exactly equivalent to Chavent et al. (2006) with the same set of solutions $\hat{\mu}^z$ and $\hat{\lambda}^z$ respectively the median of the set $\{c_z^s, s \in C\}$ of the polygon centers, and the median of the set $\{r_z^s, s \in C\}$ of their radiuses.

□

Our clustering algorithm for polygonal data can, therefore, be stated as follows (and is based on dynamic algorithm of Chavent et al., 2006):

Initialization: Define a random partition $P = (C_1, ..., C_i, ..., C_k)$

Allocation:
    test ← 0
    for $s = 1$ to $n$ do:
        Find the cluster $C_m$ to which $s$ belongs
        If $card(C_m) \neq 1$ for $l = 1, \ldots, k$ and $l \neq m$
            Perform the new prototypes $G_m$ of $C_m \backslash \{s\}$ and $G_l$ of $C_l \cup \{s\}$
            Perform the criterion $\Delta_l = \sum_{i=1}^{k} \sum_{s' \in C_i} d_H(p_{s'}, G_i)$
        Find the cluster $C_{l^*}$ such that
$$l^* = arg \min_{l=1,\ldots,k} \Delta_l$$
        If $l^* \neq m$ move $s$ to $C_{l^*}$
            test ← 1
            $C_{l^*} = C_{l^*} \cup \{s\}$ and $C_m = C_m \backslash \{s\}$
If test $= 0$ then stop, otherwise go to a

### 4. Empirical application: classification of cultural entrepreneurial regimes

For the empirical application we use data of The Global Entrepreneurship Monitor (GEM), which is a research project and annual assessment of the national level of entrepreneurial activity in multiple, diverse countries. Based in London, England, GEM is now the largest ongoing study of entrepreneurial dynamics in the world.

The data used for the GEM is collected from two large surveys, the Adult Population Survey (APS) and the National Expert Survey (NES). The APS surveys at least 2000 adults of each country covered by the GEM and covers the entrepreneurial aspirations of the country's population. The NES surveys a group of business and academic experts in each country with a broad range of specialties for concrete measures of country's institutional factors.

Each year, the GEM assembles the survey of a minimum of 2000 adults and at least 36 experts from a country of interest into an annual report. In the 2014 report, 206,000 adults from around the world anonymously participated along with 3,936 national experts.

In the application we used the following set of fifteen variables (for transformation into polygons, binary variables are transformed into their linear probabilities version):

- Gender and age of respondent;
- Q1A1 (binary). Are you, alone or with others, currently trying to start a new business;
- Q2A (binary). Are you, alone or with others, currently the owner of a business you help;
- Q4A (binary). Have you, in the past three years, personally provided funds for a new business;
- Q3A (binary). Are you, alone or with others, expecting to start a new business;
- Q3B (binary). Have you, in the past 12 months, sold, shut down, discontinued or quit a business;
- Qi1 (binary). Do you know someone personally who started a business in the past 2 years;
- Qi2 (binary). In the next six months, will there be good opportunities for starting a business;
- Qi3 (binary). Do you have the knowledge, skill and experience required to start a new business;

- Qi4 (binary). Would fear of failure would prevent you from starting a business;
- Qi5 (binary). In my country, most people would prefer that everyone had a similar standard;
- Qi6 (binary). In my country, most people consider starting a new business a desirable call;
- Qi7 (binary). In my country, those successful at starting a new business have a high level of expertise;
- Qi8 (binary). In my country, you will often see stories in the public media about success of entreprises.

We present basic descriptives for all fifteen variables in the form of polygonal variables – we use squares (4-angles) and octogons (8-angles).

**Figure 1:** Basic descriptive for 15 included variables, squares representation

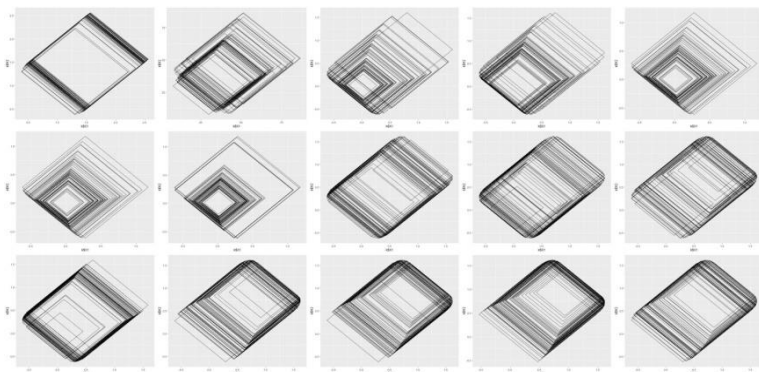

Source: Own calculations based on GEM dataset for 2014.

**Figure 2:** Basic descriptive for 15 included variables, octogons representation

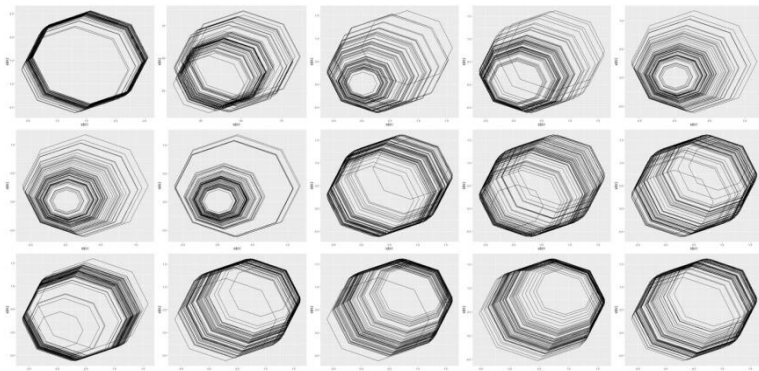

Source: Own calculations based on GEM dataset for 2014.

Results of our clustering procedure, based on both squares and octogons, are presented above.

All clustering procedures converged in reasonable time. In the best clustering solution, we got four clusters:
Cluster 1: Azores, Argentina, Bosnia and Herzegovina, Croatia, Germany, Greece, Hungary, Israel, Italy, Korea, Latvia, Macedonia, Portugal, Romania, Russia, Slovenia, Spain, Taiwan, Turkey, Uruguay
Cluster 2: Angola, Bolivia, Costa Rica, Ghana, Iran, Uganda, Vanuatu, Zambia
Cluster 3: Australia, Belgium, Finland, France, Iceland, Ireland, Japan, Malaysia, Netherlands, Norway, Sweden, Switzerland, United Kingdom, United States

Cluster 4: Brazil, Chile, China, Colombia, Ecuador, Egypt, Guatemala, Jamaica, Mexico, Montenegro, Pakistan, Peru, Saudi Arabia, Trinidad & Tobago, Tunisia, West Bank & Gaza

5. **Discussion and Conclusion**

In our article we provide several important novelties in the literature. Firstly, we derive novel and first clustering algorithm for polygonal data, which are representations of real-valued variables in $\mathbb{R}^2$ (but could be extended to $\mathbb{R}^n$ with arbitrary $n$), where the representation is defined by the polygon center and radius. This provides large methodological possibilities for work in future, as symbolic data analysis is possibly the most powerful approach to deal with big datasets, rivalled here with machine learning approaches. We also provide only second article in the literature explicitly dealing with such, polygonal type of variables.

Possibilities to extend the work in methodological terms are vast, as polygonal data analysis is only gaining ground. We could compare the results with main competitive approach in symbolic data analysis, interval data clustering. We could extend the analysis to different combinations of variables (higher or lower in number – our algorithm did not show significant problems depending upon number of included variables). Important and possibly key question is selection of polygons – the accuracy of solutions seems to grow with number of vertices chosen, but a criteria of selection, possibly based on some incremental variance ratio, would be much welcome and needed for further work. More consistent definitions of moments and regression possibilities would be necessary. Finally, extension to cross-section time-series analysis of GEM datasets (and, possibly, Amadeus) would be great, in our empirical case to validate and make the clusters robust.

For future it would be necessary to also include model based clustering algorithms (e.g. Gaussian, Dirichlet, nonparametric and other finite mixture possibilities; Bayesian and Bayesian nonparametric possibilities). It would also be great to combine the method with extension to machine learning possibilities. Crucially, it would be interesting to relax the assumption of uniform distribution within the polygon, but the same problem occurs with interval data in general. Finally, asymptotic behaviour and simulation studies should be performed to explore the performance of the procedure proposed.

Regarding possibilities of work in cultural entrepreneurship we would suggest more consideration over methodological novelties and more sophisticated methods. While the field now has a "history" and several key referential works, even regression possibilities have been largely unexplored in most aspects (e.g. non- and semiparametric methods (for example quantile regression), Bayesian modelling – which seems natural for the field, machine learning regression methods). Our article tried to provide a combination of some more complex methodological work and novelties and empirical application. We hope it will stimulate more complex methodological applications in the field.